\documentclass[aps,showkeys,showpacs,superscriptaddress]{revtex4-2}

\usepackage{amsmath,amssymb}
\usepackage[T2A]{fontenc}
\usepackage[cp1251]{inputenc}
\usepackage{epstopdf}
\usepackage[dvips]{graphicx}
\usepackage{graphicx}

\begin{document}

\title[Kinetic Model of the Emergence of Autocatalysis]%
{Kinetic Model of the Emergence of Autocatalysis}

%

\author{P.~O.~Mchedlov-Petrosyan}

\affiliation{\it National Science Center "Kharkiv Institute of Physics and Technology",   \\
1, Akademichna St., Kharkiv, 61108, Ukraine\\}

\author{L.~N.~Davydov}
\email[Corresponding author: ]{ldavydov@kipt.kharkov.ua}

\affiliation{\it National Science Center "Kharkiv Institute of Physics and Technology",   \\
1, Akademichna St., Kharkiv, 61108, Ukraine\\}


\begin{abstract}
We develop a formal model of the emergence of self-constructing objects (e.g. heteropolymers with autocatalytic capability) in an open system, which don't contain such objects initially. The objects are constructed from subunits (e.g. monomers). Each object is characterized by the difference of self-instructed reproduction and decomposition rate only. This difference, divided by a common dimensional constant, is called ``productivity''. Due to external influence the productivity of each object can randomly change. The system as a whole is subjected to external limitation: the total number of the objects is conserved (e.g., by the controlled influx of monomers). We consider such process as possibly simplest example of self-organization. We obtained exact solutions of our model for several presumed mechanisms of random change of the productivity. We have shown that the probability to find self-constructing objects in the system necessarily increases, even if initially it was equal to zero.
\end{abstract}



\maketitle

\renewcommand{\theequation}{\arabic{section}.\arabic{equation}}
\section{Introduction}\label{Sec:1}

Here we first describe our model; then we explain its origin and why it is introduced. We consider an open system that includes a manifold of chemical objects of some sort, composed of some subunits (e.g.  hetero-polymer chains composed of some monomers). We presume that among the objects of this sort there appear (even with extremely low probability) objects, capable of making self-instructed building from its subunits (monomers). We also presume that this process is a first-order reaction. Generally, particular manifold in question may not contain such self-constructing objects; it's enough to presume the possibility of formation of self-constructing objects from a given set of subunits. The difference in rate constants of self-instructed formation and decomposition, divided by a common dimensional constant $k$ will be called ``productivity''; we presume that the objects are differentiated by their productivity only. Let $x_{1} $ and $x_{2} $ to be some arbitrary productivities; each of them could be positive or negative. It is assumed that the object with productivity $x_{1} $ by a sequence (possibly a long one) of small changes could be transformed into the object with productivity $x_{2} $. Now the system is subjected to random external influences that result in changes of the objects structure and their productivity; each such impact and the corresponding change are small. Productivity change is always discrete; however, if an elementary change is small, and the system contains many objects, the system could be described by the distribution function $f\left(x,t\right)$ where the productivity $x$ is treated as a continuous variable. We presume that the total number of the objects in the system is conserved by the environmental influence (e.g., by the controlled influx of monomers), i.e.

\begin{equation} \label{1.1} \int _{-\infty }^{+\infty }f\left(x,t\right)dx=1 .  \end{equation}

Then the time evolution of $f\left(x,t\right)$ is described by the following nonlinear integral-differential equation:
\begin{eqnarray} \label{1.2}  & &{\frac{\partial f\left(x,t\right)}{\partial t} =k\left[x-\int _{-\infty }^{+\infty }y f\left(y,t\right)dy\right]f\left(x,t\right)} \nonumber \\ & &{+\int _{-\infty }^{+\infty }\left[w{\left\langle x \mathrel{\left| \vphantom{x x'}\right.\kern-\nulldelimiterspace} x' \right\rangle} f\left(x',t\right)-w{\left\langle x' \mathrel{\left| \vphantom{x' x}\right.\kern-\nulldelimiterspace} x \right\rangle} f\left(x,t\right)\right] dx} . \end{eqnarray}
Here $w{\left\langle x \mathrel{\left| \vphantom{x x'}\right.\kern-\nulldelimiterspace} x' \right\rangle} $ is the probability (per unit time) of emergence in the system of an object with the productivity $x$ due to random influence on an object with the productivity $x'$ (more precise, for a self-constructed object due to the random ``reproduction error'' of $x'$). The transition term has a standard ``master equation'' form.

This model is the continuous analogue of M.Eigen's discrete selection model \cite{1}; more precisely of its modification proposed by W.Ebeling \cite{2}. However, here we are not interested in the possible biological aspects of this model; we consider it as a physical model of probably simplest self-organization process. Our model was proposed long ago \cite{3}; however only brief description and some exact results were given; additionally, it was published in Russian and Ukrainian only. In the present work we give more detailed derivation of the results of \cite{3}, as well as discussion of some consequences and generalizations.

In \eqref{1.2} both selection and random changes of the productivity are described simultaneously within the deterministic equation (some stochastic generalizations will be considered below, see Section 4). We consider negative productivities $x<0$ as well; the non-self-constructed objects have the productivities $x\le x_{0} \le 0$, where $\left(-kx_{0} \right)$ is the minimal rate constant of such objects' disintegration.

The mechanisms of the random productivities changes are different, strictly speaking, for the domains $x\le x_{0} $ and $x>x_{0} $. However, in the next Section we consider only the case
\begin{equation} \label{1.3} w{\left\langle x \mathrel{\left| \vphantom{x x'}\right.\kern-\nulldelimiterspace} x' \right\rangle} =w\left(x-x'\right).   \end{equation}
We give the exact solution (by quadrature) of the nonlinear integral-differential master equation \eqref{1.2} for arbitrary $w\left(z\right)$ and $f\left(x,0\right)$, and discuss the behavior of mean productivity $\left\langle x\right\rangle $ and dispersion $\left\langle x^{2} \right\rangle -\left\langle x\right\rangle ^{2} $ of $f\left(x,t\right)$. In Section 3 we drop the assumption \eqref{1.3} and introduce a diffusional approximation of the master equation \eqref{1.2} which enables explicit exact solutions. In Section 4 we explore the possibility of a stochastic nature of the coefficients, arising in the diffusional approximation. In Section 5 we introduce the nonlinear autocatalysis and give several solutions for a simplified model.

\setcounter{equation}{0}
\section{General solution of equation (1.2)}\label{Sec:2}

Below we assume that both $w\left(z\right)$ and the initial distribution $f\left(x,0\right)$ are decreasing fast enough for $z,x\to \pm \infty $, so for any positive $a$ the integrals
\[\int _{-\infty }^{+\infty }\exp \left(az\right)w\left(z\right) dz;\int _{-\infty }^{+\infty }\exp \left(az\right)f\left(x,0\right) dz\]
exist. Introducing Fourier transforms,
\begin{equation} \label{2.1)} F\left(\omega ,t\right)=\frac{1}{2\pi } \int _{-\infty }^{+\infty }\exp \left(-i\omega z\right)f\left(z,t\right)dz ,   \end{equation}
\begin{equation} \label{2.2} W\left(\omega \right)=\frac{1}{2\pi } \int _{-\infty }^{+\infty }\exp \left(-i\omega z\right)w\left(z\right)dz  \end{equation}
and noticing that
\begin{equation} \label{2.3} \int _{-\infty }^{+\infty }zf\left(z,t\right)dz=2\pi i \left. \left(\frac{\partial F\left(\omega ,t\right)}{\partial \omega } \right)\right|_{\omega =0}  \end{equation}
the Fourier transformed equation \eqref{1.2} can be written as
\begin{eqnarray} \label{2.4)} & & {\frac{\partial F\left(\omega ,t\right)}{\partial t} =k\left[i\frac{\partial F\left(\omega ,t\right)}{\partial \omega } -2\pi iF\left(\omega ,t\right)\left(\left. \frac{\partial F\left(\omega ,t\right)}{\partial \omega } \right|_{\omega =0} \right)\right]} \nonumber \\ & &{+2\pi F\left(\omega ,t\right)\left[W\left(\omega \right)-W\left(0\right)\right]}. \end{eqnarray}

Substitution for $F\left(\omega ,t\right)$
\begin{equation} \label{2.5)} F\left(\omega ,t\right)=u\left(\omega ,t\right)\exp \left(-\frac{2\pi }{ik} \int _{0}^{\omega }\left[W\left(\omega '\right)-W\left(0\right)\right] d\omega '\right) \end{equation}
yields the first order equation for $u\left(\omega ,t\right)$
\begin{equation} \label{2.6)} \frac{\partial u}{\partial t} =ki\frac{\partial u}{\partial \omega } -2\pi iku\left(\left. \frac{\partial u}{\partial \omega } \right|_{\omega =0} \right) .\end{equation}

The general solution of the latter equation is
\begin{equation} \label{2.7)} u\left(\omega ,t\right)=\frac{\rho \left(\omega +ikt\right)}{1+2\pi \left[\rho \left(ikt\right)-\rho \left(0\right)\right]} , \end{equation}
where $\rho \left(y\right)$ is an arbitrary function. Correspondingly, the general solution for $F\left(\omega ,t\right)$ is
\begin{equation} \label{2.8} F\left(\omega ,t\right)=\frac{\rho \left(\omega +ikt\right)}{1+2\pi \left[\rho \left(ikt\right)-\rho \left(0\right)\right]} \exp \left(-\frac{2\pi }{ik} \int _{0}^{\omega }\left[W\left(\omega '\right)-W\left(0\right)\right] d\omega '\right) .\end{equation}

Now, using the initial distribution $f\left(x,0\right)$ we can find $\rho \left(y\right)$:
\begin{equation} \label{2.9} \rho \left(y\right)=\left[\frac{1}{2\pi } \int _{-\infty }^{+\infty }\exp \left(-iyz\right)f\left(z,0\right)dz \right]\exp \left(\frac{2\pi }{ik} \int _{0}^{y}\left[W\left(\omega '\right)-W\left(0\right)\right] d\omega '\right) .\end{equation}
Using definition of the $W\left(\omega \right)$ \eqref{2.2}, denoting for brevity $I=\int _{-\infty }^{+\infty }w\left(z\right)dz $ and changing the order of integration in the argument of exponential function in the right hand sides of \eqref{2.8} and \eqref{2.9}, we get
\begin{equation} \label{2.10)} \frac{2\pi }{ik} \int _{0}^{y}\left[W\left(\xi \right)-W\left(0\right)\right] d\xi =\frac{1}{k} \int _{-\infty }^{+\infty }\frac{w\left(z\right)}{z} \left[\exp \left(-iyz\right)-1\right]dz-I\frac{y}{ik} .  \end{equation}
Then \eqref{2.9} could be rewritten as
\begin{equation} \label{2.11}  {\rho \left(y\right)=\left[\frac{1}{2\pi } \int _{-\infty }^{+\infty }\exp \left(-iyz\right)f\left(z,0\right)dz \right] }  {\times \exp \left(\frac{1}{k} \int _{-\infty }^{+\infty }\frac{w\left(z\right)}{z} \left[\exp \left(-iyz\right)-1\right]dz-I\frac{y}{ik}  \right)} .
\end{equation}
From \eqref{1.1} and \eqref{2.11} it follows $\rho \left(0\right)=\frac{1}{2\pi } $; then \eqref{2.8} could be rewritten as
\begin{equation} \label{2.12} F\left(\omega ,t\right)=\frac{\rho \left(\omega +ikt\right)}{2\pi \rho \left(ikt\right)} \exp \left(-\frac{1}{k} \int _{-\infty }^{+\infty }\frac{w\left(z\right)}{z} \left[\exp \left(-i\omega z\right)-1\right]dz+I\frac{\omega }{ik}  \right). \end{equation}

Substitution of \eqref{2.11} for $\rho \left(\omega +ikt\right)$ and $\rho \left(ikt\right)$ into \eqref{2.12} yields the Fourier transform of the exact solution of the nonlinear integral-differential equation \eqref{1.2}, satisfying the initial condition $\left. f\right|_{t=0} =f\left(x,0\right)$. To avoid the bulk formulas we will not write down the solution by quadrature explicitly; anyway for arbitrary $w\left(x-x'\right)$ and $f\left(x,0\right)$ such expressions are practically intractable. However, some important conclusions could be drawn from the general form of the solution already. The average value of productivity is (see \eqref{2.3})
\begin{equation} \label{2.13)} \left\langle x\right\rangle =\int _{-\infty }^{+\infty }zf\left(z,t\right)dz =2\pi i\left. \left(\frac{\partial F\left(\omega ,t\right)}{\partial \omega } \right)\right|_{\omega =0}  .  \end{equation}
Using\eqref{2.12}, we obtain
\begin{equation} \label{2.14)} \left. \left(\frac{\partial F\left(\omega ,t\right)}{\partial \omega } \right)\right|_{\omega =0} =\frac{1}{2\pi } \left. \frac{d}{dy} \ln \rho \left(y\right)\right|_{y=ikt}  .  \end{equation}
Correspondingly, using\eqref{2.11} we get
\begin{equation} \label{2.15} \left\langle x\right\rangle =\frac{\int _{-\infty }^{+\infty }z\exp \left(ktz\right)f\left(z,0\right)dz }{\int _{-\infty }^{+\infty }\exp \left(ktz\right)f\left(z,0\right)dz } +\frac{1}{k} \left[\int _{-\infty }^{+\infty }w\left(z\right)\exp \left(ktz\right)dz -I\right] .  \end{equation}

This was the main result of the paper \cite{3}: even if the initial distribution lies completely in the non-self-constructing domain, i.e. $f\left(x,0\right)>0$ for $x<x_{0} <0$ only, due to the last term in the right-hand side of \eqref{2.15} the average value of productivity $\left\langle x\right\rangle $ becomes positive for large enough time. It is also interesting to consider the evolution of dispersion $\left\langle x^{2} \right\rangle -\left\langle x\right\rangle ^{2} $ of the distribution $f\left(x,t\right)$:
\begin{equation} \label{2.16} \left\langle x^{2} \right\rangle =-2\pi \left. \frac{\partial ^{2} F\left(\omega ,t\right)}{\partial \omega ^{2} } \right|_{\omega =0} =-\left\{\frac{\rho ''\left(ikt\right)}{\rho \left(ikt\right)} +\frac{1}{k} \int _{-\infty }^{+\infty }zw\left(z\right)dz \right\}.   \end{equation}
Then
\begin{equation} \label{2.17)} \left\langle x^{2} \right\rangle -\left\langle x\right\rangle ^{2} =-\left. \left[\frac{d^{2} \ln \rho \left(y\right)}{dy^{2} } \right]\right|_{y=ikt} -\frac{1}{k} \int _{-\infty }^{+\infty }zw\left(z\right)dz  .  \end{equation}

Finally we get for dispersion
\begin{eqnarray} \label{2.18} {\left\langle x^{2} \right\rangle -\left\langle x\right\rangle ^{2}} & & {=\frac{\int _{-\infty }^{+\infty }z^{2} \exp \left(zkt\right)f\left(z,0\right)dz }{\int _{-\infty }^{+\infty }\exp \left(zkt\right)f\left(z,0\right)dz } } {-\left(\frac{\int _{-\infty }^{+\infty }z\exp \left(zkt\right)f\left(z,0\right)dz }{\int _{-\infty }^{+\infty }\exp \left(zkt\right)f\left(z,0\right)dz } \right)^{2}}\nonumber \\ & & {+\frac{1}{k} \left[\int _{-\infty }^{+\infty }zw\left(z\right)\exp \left(zkt\right)dz -\int _{-\infty }^{+\infty }zw\left(z\right)dz \right]} . \end{eqnarray}

For $t=0$
\begin{equation} \label{2.19)} \left\langle x^{2} \right\rangle -\left\langle x\right\rangle ^{2} =\int _{-\infty }^{+\infty }z^{2} f\left(z,0\right)dz -\left(\int _{-\infty }^{+\infty }zf\left(z,0\right)dz \right)^{2}  .  \end{equation}
Differentiating \eqref{2.18}, we get
\begin{eqnarray} \label{2.20} \frac{d}{dt} \left(\left\langle x^{2} \right\rangle -\left\langle x\right\rangle ^{2} \right) & &=k\frac{\int _{-\infty }^{+\infty }z^{3} \exp \left(zkt\right)f\left(z,0\right)dz }{\int _{-\infty }^{+\infty }\exp \left(zkt\right)f\left(z,0\right)dz }  \nonumber \\ & &{-3k\frac{\int _{-\infty }^{+\infty }z^{2} \exp \left(zkt\right)f\left(z,0\right)dz\int _{-\infty }^{+\infty }z\exp \left(zkt\right)f\left(z,0\right)dz  }{\left(\int _{-\infty }^{+\infty }\exp \left(zkt\right)f\left(z,0\right)dz \right)^{2} } } \nonumber \\ & &{+2k\left(\frac{\int _{-\infty }^{+\infty }z\exp \left(zkt\right)f\left(z,0\right)dz }{\int _{-\infty }^{+\infty }\exp \left(zkt\right)f\left(z,0\right)dz } \right)^{3} +\int _{-\infty }^{+\infty }z^{2} w\left(z\right)\exp \left(zkt\right)dz } . \end{eqnarray}

Evidently, for large enough times the last positive term will dominate, and the dispersion will increase. Thus, while the transition from non-self-constructing to self-constructing objects is inevitable, asymptotically there is no ``selection'', i.e. the distribution function $f\left(x,t\right)$ flattens asymptotically.

It is illustrative to consider the Gaussian forms both for transition probability $w\left(z\right)$ and an initial distribution $f\left(z,0\right)$ as an example:
\begin{equation} \label{2.21} w\left(z\right)=\mu \frac{\beta }{\sqrt{\pi } } \exp \left[-\beta ^{2} z^{2} \right] ,  \end{equation}
\begin{equation} \label{2.22} f\left(x,0\right)=\frac{\alpha }{\sqrt{\pi } } \exp \left[-\alpha ^{2} \left(x-X\right)^{2} \right]\, .  \end{equation}
Here $\mu $ is a constant of the same dimensionality as $k$; also we note that for $\alpha \to \infty$ and $\beta \to \infty $ both \eqref{2.21} and \eqref{2.22} approach $\delta $-functions.

Then we obtain from \eqref{2.15} for the average productivity
\begin{equation} \label{2.23} \left\langle x\right\rangle =X+\frac{kt}{2\alpha ^{2} } +\frac{\mu }{k} \left[\exp \left(\frac{kt}{2\beta } \right)^{2} -1\right].   \end{equation}
Similarly from \eqref{2.18} we get for the dispersion:
\begin{equation} \label{2.24} \left\langle x^{2} \right\rangle -\left\langle x\right\rangle ^{2} =\frac{1}{2\alpha ^{2} } +\mu \exp \left[\left(\frac{kt}{2\beta } \right)^{2} \right]\left[\frac{t}{2\beta ^{2} } \right] .  \end{equation}

\section{Diffusional approximation for the master equation}\label{Sec:3}
\setcounter{equation}{0}

Starting again from the master equation \eqref{1.2} we consider the transition term $T$:
\begin{equation} \label{3.1)} T=\int _{-\infty }^{+\infty }\left[w{\left\langle x \mathrel{\left| \vphantom{x x'}\right.\kern-\nulldelimiterspace} x' \right\rangle} f\left(x',t\right)-w{\left\langle x' \mathrel{\left| \vphantom{x' x}\right.\kern-\nulldelimiterspace} x \right\rangle} f\left(x,t\right)\right] dx .  \end{equation}
However, we drop our previous assumption \eqref{1.3}. Following, e.g. Van Kampen \cite{4}, we express the transition probability as a function of the length of the jump $r$ and the starting point,
\begin{equation} \label{3.2} w{\left\langle x \mathrel{\left| \vphantom{x x'}\right.\kern-\nulldelimiterspace} x' \right\rangle} =w\left(x',r\right);\, \, \, \, \, r=x-x'.   \end{equation}
Then the transition term $T$ reads
\begin{equation} \label{3.3} T=\int _{-\infty }^{+\infty }w\left(x-r,r\right)f\left(x-r,t\right) dr-f\left(x,t\right)\int _{-\infty }^{+\infty }w\left(x,-r\right) dr .  \end{equation}

The basic assumption is that $w\left(x',r\right)$ is a sharply peaked function of $r$ but varies slowly with $x'$. If $f\left(x,t\right)$ also varies slowly with $x$ on the scale of (fast) variation of $w$ with $r$, we may expand the integrand in a Taylor series, keeping the terms up to second order:
\begin{eqnarray} \label{3.4)} 
w\left(x-r,r\right)f\left(x-r,t\right)& &=w\left(x,r\right)f\left(x,t\right)- \frac{\partial }{\partial x} \left. \left[w\left(x-r,r\right)f\left(x-r,t\right)\right]\right|_{r=0} \cdot r \nonumber \\ & &{+\frac{1}{2} \frac{\partial ^{2} }{\partial x^{2} } \left. \left[w\left(x-r,r\right)f\left(x-r,t\right)\right]\right|_{r=0} \cdot r^{2} -...}
\end{eqnarray}
Then $T$ \eqref{3.3} takes the form
\begin{eqnarray} \label{3.5)} 
T& &=f\left(x,t\right)\int _{-\infty }^{+\infty }w\left(x,r\right) dr-\int _{-\infty }^{+\infty }r\frac{\partial }{\partial x} \left[w\left(x,r\right)f\left(x,t\right)\right] dr \nonumber \\ & &{+\frac{1}{2} \int _{-\infty }^{+\infty }r^{2} \frac{\partial ^{2} }{\partial x^{2} } \left[w\left(x,r\right)f\left(x,t\right)\right] dr-f\left(x,t\right)\int _{-\infty }^{+\infty }w\left(x,-r\right) dr} . \end{eqnarray}
Changing the order of differentiation in $x$ and integrating over $r$, we obtain
\begin{equation} \label{3.6)} T=-\frac{\partial }{\partial x} \left[f\left(x,t\right)\int _{-\infty }^{+\infty }rw\left(x,r\right) dr\right]+\frac{1}{2} \frac{\partial ^{2} }{\partial x^{2} } \left[f\left(x,t\right)\int _{-\infty }^{+\infty }r^{2} w\left(x,r\right) dr\right] .\end{equation}
Introducing the moments of the jump probability
\begin{equation} \label{3.7} \phi \left(x\right)=\int _{-\infty }^{+\infty }rw\left(x,r\right) dr;\, \, \, \, \, \, \psi \left(x\right)=\int _{-\infty }^{+\infty }r^{2} w\left(x,r\right) dr \end{equation}
we rewrite \eqref{1.2} as
\begin{eqnarray} \label{3.8}  \frac{\partial f\left(x,t\right)}{\partial t} & &=k\left[x-\int _{-\infty }^{+\infty }y f\left(y,t\right)dy\right]f\left(x,t\right)\nonumber \\ & &{-\frac{\partial }{\partial x} \left[\phi \left(x\right)f\left(x,t\right)\right]+\frac{1}{2} \frac{\partial ^{2} }{\partial x^{2} } \left[\psi \left(x\right)f\left(x,t\right)\right]} . \end{eqnarray}

In this Section we will neglect the $x$- dependence of $\phi $ and $\psi $, setting
\begin{equation} \label{3.9)} \phi \left(x\right)=\nu =\textrm{const};\, \, \, \, \frac{1}{2} \psi \left(x\right)=\zeta =\textrm{const}. \end{equation}
Introducing for convenience
\begin{equation} \label{3.10} \gamma \left(t\right)=\int _{-\infty }^{+\infty }zf\left(z,t\right)dz \, \, , \end{equation}
we finally rewrite \eqref{3.8} as
\begin{equation} \label{3.11} \frac{\partial f\left(x,t\right)}{\partial t} =k\left[x-\gamma \left(t\right)\right]f\left(x,t\right)-\nu \frac{\partial f\left(x,t\right)}{\partial x} +\zeta \frac{\partial ^{2} f\left(x,t\right)}{\partial x^{2} } . \end{equation}

Performing the Fourier transform, we get
\begin{equation} \label{3.12)} \frac{\partial F\left(\omega ,t\right)}{\partial t} =ik\frac{\partial F\left(\omega ,t\right)}{\partial \omega } -\left[k\gamma \left(t\right)+i\nu \omega +\zeta \omega ^{2} \right]F\left(\omega ,t\right) .  \end{equation}
Substitution for $F\left(\omega ,t\right)$
\begin{eqnarray} \label{3.13}  & &{F\left(\omega ,t\right)=u\left(\omega ,t\right)\exp \left[-k\Gamma \left(t\right)+\frac{1}{2k} \nu \omega ^{2} +\frac{1}{3ki} \zeta \omega ^{3} \right]}, \nonumber \\ & &{\Gamma \left(t\right)=\int _{0}^{t}\gamma \left(t'\right)dt' },  
\end{eqnarray}
yields
\begin{equation} \label{3.14)} \frac{\partial u\left(\omega ,t\right)}{\partial t} =ik\frac{\partial u\left(\omega ,t\right)}{\partial \omega } . \end{equation}
The solution is $u=\rho \left(\omega +ikt\right)$, where $\rho \left(y\right)$ is an arbitrary function. Now, from this form of solution and \eqref{3.13} it follows
\begin{equation} \label{3.15} \left. \frac{\partial F\left(\omega ,t\right)}{\partial \omega } \right|_{\omega =0} =\left. \frac{d\rho \left(y\right)}{dy} \right|_{y=ikt} \times \exp \left[-k\Gamma \left(t\right)\right] .  \end{equation}

On the other hand, using \eqref{2.3} and the definition of the $\gamma \left(t\right)$ \eqref{3.10}, we get
\begin{equation} \label{3.16)} \left. 2\pi i\frac{\partial F\left(\omega ,t\right)}{\partial \omega } \right|_{\omega =0} =\gamma \left(t\right) .  \end{equation}
Denoting $R\left(t\right)=\exp \left[k\Gamma \left(t\right)\right]$ and $R\left(0\right)=1$, we obtain equation for finding $R$:
\begin{equation} \label{3.17} \frac{1}{ik} \frac{dR}{dt} =2\pi \left. \frac{d\rho \left(y\right)}{dy} \right|_{y=ikt}  .  \end{equation}
I.e. $R=\exp \left[\Gamma \left(t\right)\right]=1-2\pi \rho \left(0\right)+2\pi \rho \left(ikt\right)$, and we rewrite \eqref{3.13} as
\begin{equation} \label{3.18)} F\left(\omega ,t\right)=\frac{\rho \left(\omega +ikt\right)}{1-2\pi \rho \left(0\right)+2\pi \rho \left(ikt\right)} \exp \left[\frac{1}{2k} \nu \omega ^{2} -\frac{i}{3k} \zeta \omega ^{3} \right] .  \end{equation}
Using the initial condition $\left. f\left(x,t\right)\right|_{t=0} =f\left(x,0\right)$ and introducing
\begin{equation} \label{3.19} F_{0} \left(\omega \right)=\left. F\left(\omega ,t\right)\right|_{t=0} =\frac{1}{2\pi } \int _{-\infty }^{+\infty }\exp \left(-i\omega z\right)f\left(z,0\right)dz \, \, \, , \end{equation}
we get
\begin{equation} \label{3.20)} \rho \left(y\right)=F_{0} \left(y\right)\exp \left[-\frac{1}{2k} \nu y^{2} +\frac{i}{3k} \zeta y^{3} \right] .  \end{equation}
due to condition \eqref{1.1} $\rho \left(0\right)=F_{0} \left(0\right)=\frac{1}{2\pi } $. Then we obtain finally the Fourier transform of solution of the equation \eqref{3.11}
\begin{equation} \label{3.21} F\left(\omega ,t\right)=\frac{F_{0} \left(\omega +ikt\right)}{2\pi F_{0} \left(ikt\right)} \exp \left[-i\omega \left(\nu t+\zeta kt^{2} \right)-\zeta \omega ^{2} t\right] .  \end{equation}
The average value of productivity is now
\begin{equation} \label{3.22} \left\langle x\right\rangle =2\pi i\left. \left(\frac{\partial F\left(\omega ,t\right)}{\partial \omega } \right)\right|_{\omega =0} =\frac{\int _{-\infty }^{+\infty }z\exp \left(zkt\right)f\left(z,0\right)dz }{\int _{-\infty }^{+\infty }\exp \left(zkt\right)f\left(z,0\right)dz } +\left(\nu t+\zeta kt^{2} \right) .  \end{equation}

Interestingly, for the ``negative drift'' $\nu <0$, i.e. asymmetric dependence of $w\left(x,r\right)$ on $r$, the increase of $\left\langle x\right\rangle $, while asymptotically inevitable, could be non-monotonous.
To calculate dispersion, we use \eqref{2.16}
\begin{eqnarray} \label{3.23)}  \left\langle x^{2} \right\rangle & &=-2\pi \left. \frac{\partial ^{2} F\left(\omega ,t\right)}{\partial \omega ^{2} } \right|_{\omega =0} \nonumber \\ & &{=\frac{\int _{-\infty }^{+\infty }z^{2} \exp \left(zkt\right)f\left(z,0\right)dz }{\int _{-\infty }^{+\infty }\exp \left(zkt\right)f\left(z,0\right)dz } +2\frac{\int _{-\infty }^{+\infty }z\exp \left(zkt\right)f\left(z,0\right)dz }{\int _{-\infty }^{+\infty }\exp \left(zkt\right)f\left(z,0\right)dz } \times \left(\nu t+\zeta kt^{2} \right)+} \left(\nu t+\zeta kt^{2} \right)^{2} +2\zeta t . \end{eqnarray}
Correspondingly, dispersion is
\begin{equation} \label{3.24} \left\langle x^{2} \right\rangle -\left\langle x\right\rangle ^{2} =\frac{\int _{-\infty }^{+\infty }z^{2} \exp \left(zkt\right)f\left(z,0\right)dz }{\int _{-\infty }^{+\infty }\exp \left(zkt\right)f\left(z,0\right)dz } -\left[\frac{\int _{-\infty }^{+\infty }z\exp \left(zkt\right)f\left(z,0\right)dz }{\int _{-\infty }^{+\infty }\exp \left(zkt\right)f\left(z,0\right)dz } \right]^{2} +2\zeta t .  \end{equation}
I.e. while for intermediate times the dispersion is influenced by kinetics, asymptotically it is increasing linear in time, i.e. determined by the diffusional transfer only.

Let us consider again the Gaussian initial distribution, see \eqref{2.22}, centered on some productivity $X$:
\begin{equation} \label{3.25} f\left(x,0\right)=\frac{\alpha }{\sqrt{\pi } } \exp \left[-\alpha ^{2} \left(x-X\right)^{2} \right] .  \end{equation}
Using \eqref{3.22} we get explicit expression of the average productivity for this initial distribution
\begin{equation} \label{3.26)} \left\langle x\right\rangle =X+\left(\frac{k}{2\alpha ^{2} } +\nu \right)t+\zeta kt^{2}  .  \end{equation}
If there is $\nu <0$, i.e. the ``backwards jumps'' are preferable,
\begin{equation} \label{3.27)} \frac{d\left\langle x\right\rangle }{dt} =\frac{k}{2\alpha ^{2} } -\left|\nu \right|+2\zeta kt .  \end{equation}
This means, that if $0<\left|\nu \right|-\frac{k}{2\alpha ^{2} } $, the average productivity will decrease till the moment
\begin{equation} \label{3.28} t_{s} =\frac{1}{2\zeta k} \left(\left|\nu \right|-\frac{k}{2\alpha ^{2} } \right)=\frac{1}{\zeta } \left(\frac{\left|\nu \right|}{2k} -\frac{1}{4\alpha ^{2} } \right) .  \end{equation}

Evidently, $t_{s} $ decreases with increase of ``diffusivity'' and with the increase of the ratio of reaction rate to the asymmetry of jump probability. And $t_{s} $ increases with the ``narrowness'' of the initial distribution; for $\alpha \to \infty $ the initial distribution given by \eqref{3.25} approaches the $\delta $-function.
Using expression \eqref{3.24} for the dispersion, we get
\begin{equation} \label{3.29)} \left\langle x^{2} \right\rangle -\left\langle x\right\rangle ^{2} =\frac{1}{2\alpha ^{2} } +2\zeta t .  \end{equation}
So for the initial distribution \eqref{3.25} the dispersion will increase monotonically.
Finally, performing the inverse Fourier transform of \eqref{3.21} we obtain the exact solution of equation \eqref{3.11}
\begin{eqnarray} \label{3.30}  f\left(x,t\right)& &=\left[2\sqrt{\pi \zeta t} \int _{-\infty }^{+\infty }\exp \left[ktz\right]f\left(z,0\right)dz \right]^{-1}  \nonumber \\ & &{\times \int _{-\infty }^{+\infty }f\left(z,0\right)\exp \left[ktz-\frac{\left(x-z-\nu t-\zeta kt^{2} \right)^{2} }{4\zeta t} \right]dz } . \end{eqnarray}
For the initial distribution $f\left(x,0\right)=\delta \left(x-X\right)$ \eqref{3.30} simplifies to
\begin{equation} \label{3.31} f\left(x,t\right)=\frac{1}{2\sqrt{\pi \zeta t} } \exp \left[-\frac{\left(x-X-\nu t-\zeta kt^{2} \right)^{2} }{4\zeta t} \right] .  \end{equation}
This solution looks very like a solution of a diffusion equation with a drift, however with a fast accelerating drift.

\section{Diffusional approximation with random diffusion and drift}\label{Sec:4}
\setcounter{equation}{0}

Considering the master equation in the previous Section we have, following \cite{4}, neglected the dependence of the jump probability moments \eqref{3.7} on $x$, i.e. on  the starting point of the jump. Another reasonable assumption could be: to presume this dependence to be a random function of $x$. Starting again with equation \eqref{3.8}, we presume now that
\begin{equation} \label{4.1} \phi \left(x\right)=\nu +\sigma \eta \left(x\right),\, \, \, \, \frac{1}{2} \psi \left(x\right)=\zeta +s\eta \left(x\right), \end{equation}
where $\eta \left(x\right)$ is a Gaussian random function, $\left\langle \eta \left(x\right)\right\rangle =0$, and the correlator is
\begin{equation} \label{4.2} \left\langle \eta \left(x\right)\eta \left(x'\right)\right\rangle =q\left(x-x'\right) .  \end{equation}
To avoid confusion we note that in the present Section the parenthesis $\left\langle \right\rangle $ will denote averaging over $\eta \left(x\right)$, and the averaging over the distribution $f\left(x,t\right)$ will be denoted as $\left\langle \right\rangle _{f} $. In \eqref{4.1} $\sigma $ and $s$ are some constants, depending on the particular functional form of $w\left(x,r\right)$, see \eqref{3.2} and \eqref{3.7}. Then equation \eqref{3.8} takes the form
\begin{equation} \label{4.3}  {\frac{\partial f\left(x,t\right)}{\partial t} =k\left[x-\gamma \left(t\right)\right]f\left(x,t\right)} {-\frac{\partial }{\partial x} \left\{\left[\nu +\sigma \eta \left(x\right)\right]f\left(x,t\right)\right\}+\frac{\partial ^{2} }{\partial x^{2} } \left\{\left[\zeta +s\eta \left(x\right)\right]f\left(x,t\right)\right\}} . \end{equation}
To avoid superfluous complications we define the analog of the condition \eqref{1.1} as
\begin{equation} \label{4.4} \int _{-\infty }^{+\infty }\left\langle f\left(x,t\right)\right\rangle dx=1  .  \end{equation}
Correspondingly, it is convenient to define
\begin{equation} \label{4.5)} \gamma \left(t\right)=\int _{-\infty }^{+\infty }x\left\langle f\left(x,t\right)\right\rangle dx  .  \end{equation}
Formally averaging over fluctuations of $\eta \left(x\right)$ yields
\begin{eqnarray} \label{4.6}  \frac{\partial \left\langle f\left(x,t\right)\right\rangle }{\partial t} & &=k\left[x-\gamma \left(t\right)\right]\left\langle f\left(x,t\right)\right\rangle \nonumber \\ & &{-\frac{\partial }{\partial x} \left[\nu \left\langle f\left(x,t\right)\right\rangle +\sigma \left\langle \eta \left(x\right)f\left(x,t\right)\right\rangle \right]+\frac{\partial ^{2} }{\partial x^{2} } \left[\zeta \left\langle f\left(x,t\right)\right\rangle +s\left\langle \eta \left(x\right)f\left(x,t\right)\right\rangle \right]} . \end{eqnarray}

To obtain the correlator $\left\langle \eta \left(x\right)f\left(x,t\right)\right\rangle $ we use the Novikov-Furutsu formula \cite{5}:
\begin{equation} \label{4.7} \left\langle \eta \left(x\right)f\left(x,t\right)\right\rangle =\int _{-\infty }^{+\infty }\left\langle \eta \left(x\right)\eta \left(x'\right)\right\rangle \left\langle \frac{\delta f\left(x,t\right)}{\delta \eta \left(x'\right)} \right\rangle dx' .   \end{equation}
The formal solution of the equation \eqref{4.3} can be written as
\begin{equation} \label{4.8)}  {f\left(x,t\right)}  {=\exp \left\{t\left[kx-\frac{\partial }{\partial x} \left[\nu +\sigma \eta \left(x\right)\right]+\frac{\partial ^{2} }{\partial x^{2} } \left[\zeta +s\eta \left(x\right)\right]\right]-k\int _{0}^{t}\gamma \left(\tau \right)d\tau  \right\}f\left(x,0\right)} . \end{equation}
Here the action of the exponential operator on $f\left(x,0\right)$ is understood as the action of the Taylor expansion of the exponential function. Then
\begin{equation} \label{4.9)} \frac{\delta f\left(x,t\right)}{\delta \eta \left(x'\right)} =t\left[-\sigma \frac{\partial }{\partial x} \left[\frac{\delta \eta \left(x\right)}{\delta \eta \left(x'\right)} \right]+s\frac{\partial ^{2} }{\partial x^{2} } \left[\frac{\delta \eta \left(x\right)}{\delta \eta \left(x'\right)} \right]\right]f\left(x,t\right) .  \end{equation}
Averaging the latter expression we get
\begin{equation} \label{4.10)} \left\langle \frac{\delta f\left(x,t\right)}{\delta \eta \left(x'\right)} \right\rangle =t\left[-\sigma \frac{\partial }{\partial x} \left[\delta \left(x-x'\right)\left\langle f\left(x,t\right)\right\rangle \right]+s\frac{\partial ^{2} }{\partial x^{2} } \left[\delta \left(x-x'\right)\left\langle f\left(x,t\right)\right\rangle \right]\right].   \end{equation}
Taking the latter expression and expression \eqref{4.2} for the correlator, inserting them into \eqref{4.7}, we obtain
\begin{eqnarray} \label{4.11)}  {\left\langle \eta \left(x\right)f\left(x,t\right)\right\rangle }  & &{=-\sigma t\int _{-\infty }^{+\infty }q\left(x-x'\right)\left[\frac{\partial }{\partial x} \left[\delta \left(x-x'\right)\left\langle f\left(x,t\right)\right\rangle \right]\right]dx' }\nonumber \\ & &{+st\int _{-\infty }^{+\infty }q\left(x-x'\right)\left[\frac{\partial ^{2} }{\partial x^{2} } \left[\delta \left(x-x'\right)\left\langle f\left(x,t\right)\right\rangle \right]\right]dx' } .\end{eqnarray}
After differentiating, integrating by parts and rearranging terms we obtain
\begin{eqnarray} \label{4.12}  \left\langle \eta \left(x\right)f\left(x,t\right)\right\rangle & &=st\left[q\left(0\right)\frac{\partial ^{2} }{\partial x^{2} } \left\langle f\left(x,t\right)\right\rangle \right]-\sigma t\left[q\left(0\right)\frac{\partial }{\partial x} \left\langle f\left(x,t\right)\right\rangle \right] \nonumber \\ & &{+\left[\sigma \left\langle f\left(x,t\right)\right\rangle -2s\frac{\partial }{\partial x} \left\langle f\left(x,t\right)\right\rangle \right]tq'\left(0\right)+stq''\left(0\right)\left\langle f\left(x,t\right)\right\rangle } . \end{eqnarray}

We presume the correlator $q\left(z\right)$ to be a smooth differentiable function, therefore $q'\left(0\right)=0$. Substitution of \eqref{4.12} for $\left\langle \eta \left(x\right)f\left(x,t\right)\right\rangle $ into \eqref{4.6} yields
\begin{eqnarray} \label{4.13}  \frac{\partial \left\langle f\left(x,t\right)\right\rangle }{\partial t} & &=k\left[x-\gamma \left(t\right)\right]\left\langle f\left(x,t\right)\right\rangle -\left[\nu +\sigma stq''\left(0\right)\right]\frac{\partial }{\partial x} \left\langle f\left(x,t\right)\right\rangle \nonumber \\ & &{+\left[\zeta +\sigma ^{2} tq\left(0\right)+s^{2} tq''\left(0\right)\right]\frac{\partial ^{2} }{\partial x^{2} } \left\langle f\left(x,t\right)\right\rangle }\nonumber \\ & &{-2s\sigma tq\left(0\right)\frac{\partial ^{3} }{\partial x^{3} } \left\langle f\left(x,t\right)\right\rangle +s^{2} tq\left(0\right)\frac{\partial ^{4} }{\partial x^{4} } \left\langle f\left(x,t\right)\right\rangle } . \end{eqnarray}
It is convenient to introduce notations
\begin{equation} \label{4.14)} \theta _{1} =\sigma sq''\left(0\right);\, \, \, \theta _{2} =\sigma ^{2} q\left(0\right)+s^{2} q''\left(0\right);\, \, \, \theta _{3} =2s\sigma q\left(0\right);\, \, \, \theta _{4} =s^{2} q\left(0\right).   \end{equation}
Then we can rewrite \eqref{4.13} in a more compact form:
\begin{eqnarray} 
\label{4.15}  \frac{\partial \left\langle f\left(x,t\right)\right\rangle }{\partial t} & &=k\left[x-\gamma \left(t\right)\right]\left\langle f\left(x,t\right)\right\rangle -\left[\nu +\theta _{1} t\right]\frac{\partial }{\partial x} \left\langle f\left(x,t\right)\right\rangle \nonumber \\ & &{+\left[\zeta +\theta _{2} t\right]\frac{\partial ^{2} }{\partial x^{2} } \left\langle f\left(x,t\right)\right\rangle -\theta _{3} t\frac{\partial ^{3} }{\partial x^{3} } \left\langle f\left(x,t\right)\right\rangle +\theta _{4} t\frac{\partial ^{4} }{\partial x^{4} } \left\langle f\left(x,t\right)\right\rangle }.
\end{eqnarray}
Introducing the Fourier transform,
\begin{equation} \label{4.16)} F\left(\omega ,t\right)=\frac{1}{2\pi } \int _{-\infty }^{+\infty }\exp \left(-i\omega z\right)\left\langle f\left(z,t\right)\right\rangle dz  ,  \end{equation}
we obtain from \eqref{4.15}
\begin{equation} \label{4.17)}  \frac{\partial F\left(\omega ,t\right)}{\partial t} =ik\frac{\partial }{\partial \omega } F\left(\omega ,t\right) {-\left\{k\gamma \left(t\right)+i\nu \omega +\zeta \omega ^{2} +t\left[i\theta _{1} \omega +\theta _{2} \omega ^{2} -i\theta _{3} \omega ^{3} -\theta _{4} \omega ^{4} \right]\right\}F\left(\omega ,t\right)} . \end{equation}
Denoting
\begin{equation} \label{4.18)} \Omega \left(\omega \right)=i\theta _{1} \omega +\theta _{2} \omega ^{2} -i\theta _{3} \omega ^{3} -\theta _{4} \omega ^{4}    \end{equation}
and substituting for $F\left(\omega ,t\right)$ the following expression
\begin{eqnarray} \label{4.19}  & &{F\left(\omega ,t\right)=u\left(\omega ,t\right)\exp \left[-k\Gamma \left(t\right)+\frac{1}{2k} \nu \omega ^{2} +\frac{1}{3ki} \zeta \omega ^{3} \right]}, \nonumber \\ & &{\Gamma \left(t\right)=\int _{0}^{t}\gamma \left(t'\right)dt' }  \end{eqnarray}
we obtain the following equation for $u\left(\omega ,t\right)$
\begin{equation} \label{4.20} \frac{\partial u\left(\omega ,t\right)}{\partial t} =ik\frac{\partial }{\partial \omega } u\left(\omega ,t\right)-t\Omega \left(\omega \right)u\left(\omega ,t\right) .  \end{equation}

The corresponding system is
\begin{equation} \label{4.21)} \frac{dt}{1} =-\frac{d\omega }{ik} =-\frac{du\left(\omega ,t\right)}{t\Omega \left(\omega \right)u\left(\omega ,t\right)}  .  \end{equation}
The first integral is $C_{1} =\omega +ikt$; then $t=\frac{i}{k} \left(\omega -C_{1} \right)$, and
\begin{equation} \label{4.22)} \frac{1}{k^{2} } \left(\omega -C_{1} \right)\Omega \left(\omega \right)d\omega =\frac{du}{u}  .  \end{equation}
Integrating, we obtain the second integral
\begin{equation} \label{4.23)} C_{2} =u\exp \left\{-\frac{1}{k^{2} } \int _{}^{\omega }\left(\omega '-C_{1} \right)\Omega \left(\omega '\right)d\omega ' \right\} .  \end{equation}
Performing integration of the exponent, we obtain
\begin{eqnarray} \label{4.24)}  C_{2} & &=u\exp \left\{-\frac{1}{k^{2} } \left[\frac{i\theta _{1} }{3} \omega ^{3} +\frac{\theta _{2} }{4} \omega ^{4} -\frac{i\theta _{3} }{5} \omega ^{5} -\frac{\theta _{4} }{6} \omega ^{6} \right. \right. \nonumber \\ & &{-\left. \left. C_{1} \left(\frac{i\theta _{1} }{2} \omega ^{2} +\frac{\theta _{2} }{3} \omega ^{3} -\frac{i\theta _{3} }{4} \omega ^{4} -\frac{\theta _{4} }{5} \omega ^{5} \right)\right]\right\}} . \end{eqnarray}

The general solution of \eqref{4.20} is $P\left(C_{1} ,C_{2} \right)=0$, where $P$ is an arbitrary function of its arguments. Resolving for $C_{2} $ we get
\begin{eqnarray} \label{4.25}  u & &=p\left(\omega +ikt\right)\exp \left\{\frac{1}{k^{2} } \left[\frac{i\theta _{1} }{3} \omega ^{3} +\frac{\theta _{2} }{4} \omega ^{4} -\frac{i\theta _{3} }{5} \omega ^{5} -\frac{\theta _{4} }{6} \omega ^{6} \right. \right. \nonumber \\ & &{-\left. \left. \left(\omega +ikt\right)\left(\frac{i\theta _{1} }{2} \omega ^{2} +\frac{\theta _{2} }{3} \omega ^{3} -\frac{i\theta _{3} }{4} \omega ^{4} -\frac{\theta _{4} }{5} \omega ^{5} \right)\right]\right\}} . \end{eqnarray}
Here $p\left(y\right)$ is an arbitrary function. Rearranging terms, and denoting
\begin{eqnarray} \label{4.26)}  Q\left(\omega ,t\right)& &={\left\{\frac{1}{k^{2} } \left[\frac{\theta _{1} kt}{2} \omega ^{2} +\left(-\frac{i\theta _{1} }{6} -\frac{\theta _{2} ikt}{3} \right)\right. \right. \omega ^{3} +\left(-\frac{\theta _{2} }{12} -\frac{\theta _{3} kt}{4} \right)\omega ^{4} } \nonumber \\ & &{+\left(\frac{i\theta _{3} }{20} +\frac{\theta _{4} ikt}{5} \right)\omega ^{5} +\left. \left. \left(\frac{\theta _{4} }{30} \right)\omega ^{6} \right]\right\}}.  \end{eqnarray}
we rewrite \eqref{4.25} as
\begin{equation} \label{4.27)} u=p\left(\omega +ikt\right)\exp \left[Q\left(\omega ,t\right)\right] .  \end{equation}
With the latter expression for $u$ Eq.\eqref{4.19} becomes
\begin{equation} \label{4.28} F\left(\omega ,t\right)=p\left(\omega +ikt\right)\exp \left[-k\Gamma \left(t\right)+\frac{1}{2k} \nu \omega ^{2} +\frac{1}{3ki} \zeta \omega ^{3} +Q\left(\omega ,t\right)\right].   \end{equation}
Similar to \eqref{2.3}, we have, using the definition of $\gamma \left(t\right)$,
\begin{equation} \label{4.29)} \gamma \left(t\right)=\int _{-\infty }^{+\infty }z\left\langle f\left(z,t\right)\right\rangle dz=2\pi i \left. \left(\frac{\partial F\left(\omega ,t\right)}{\partial \omega } \right)\right|_{\omega =0}  .  \end{equation}
Differentiation of \eqref{4.28} yields
\begin{equation} \label{4.30)} \left. \frac{\partial F\left(\omega ,t\right)}{\partial \omega } \right|_{\omega =0} =\left. \frac{dp\left(y\right)}{dy} \right|_{y=ikt} \times \exp \left[-k\Gamma \left(t\right)\right] .  \end{equation}

Proceeding in the same way, as above, see \eqref{3.15}-\eqref{3.17}, we again find $\exp \left[\Gamma \left(t\right)\right]=1-2\pi p\left(0\right)+2\pi p\left(ikt\right)$; and \eqref{4.28} takes the form
\begin{equation} \label{4.31)}  F\left(\omega ,t\right)=\frac{p\left(\omega +ikt\right)}{1-2\pi p\left(0\right)+2\pi p\left(ikt\right)}  {\times \exp \left[\frac{1}{2k} \nu \omega ^{2} +\frac{1}{3ki} \zeta \omega ^{3} +Q\left(\omega ,t\right)\right]} . \end{equation}
Introducing $Q_{0} \left(\omega \right)=\left. Q\left(\omega ,t\right)\right|_{t=0} $
\begin{equation} \label{4.32)} Q_{0} \left(\omega \right)=\left. Q\left(\omega ,t\right)\right|_{t=0} =\left\{\frac{1}{k^{2} } \left[-\frac{i\theta _{1} }{6} \omega ^{3} -\frac{\theta _{2} }{12} \omega ^{4} +\frac{i\theta _{3} }{20} \omega ^{5} +\frac{\theta _{4} }{30} \omega ^{6} \right]\right\} \end{equation}
and using $F_{0} \left(\omega \right)=\left. F\left(\omega ,t\right)\right|_{t=0} $, we can find $p\left(y\right)$:
\begin{equation} \label{4.33)} p\left(y\right)=F_{0} \left(y\right)\exp \left[-\frac{1}{2k} \nu y^{2} +\frac{i}{3k} \zeta y^{3} -Q_{0} \left(y\right)\right] .  \end{equation}
Due to the condition \eqref{4.4} and $F_{0} \left(0\right)=\frac{1}{2\pi } $, there is $p\left(0\right)=\frac{1}{2\pi } $. Denoting for brevity
\begin{equation} \label{4.34} G\left(\omega ,t\right)=Q\left(\omega ,t\right)+Q_{0} \left(ikt\right)-Q_{0} \left(\omega +ikt\right), \end{equation}
we finally obtain the Fourier transform of the exact solution of equation \eqref{4.15}
\begin{equation} \label{4.35)} F\left(\omega ,t\right)=\frac{F_{0} \left(\omega +ikt\right)}{2\pi F_{0} \left(ikt\right)} \exp \left\{-i\omega \left(\nu t+\zeta kt^{2} \right)-\zeta \omega ^{2} t+G\left(\omega ,t\right)\right\}, \end{equation}
\begin{equation} \label{4.36)}  {Q\left(\omega ,t\right)=\left\{\frac{1}{k^{2} } \left[\frac{\theta _{1} kt}{2} \omega ^{2} +\left(-\frac{i\theta _{1} }{6} -\frac{\theta _{2} ikt}{3} \right)\right. \right. \omega ^{3} +\left(-\frac{\theta _{2} }{12} -\frac{\theta _{3} kt}{4} \right)\omega ^{4} }  {+\left(\frac{i\theta _{3} }{20} +\frac{\theta _{4} ikt}{5} \right)\omega ^{5} +\left. \left. \left(\frac{\theta _{4} }{30} \right)\omega ^{6} \right]\right\}} , \end{equation}
\begin{equation} \label{4.37)}  {Q_{0} \left(\omega +ikt\right)} {=\left\{\frac{1}{k^{2} } \left[-\frac{i\theta _{1} }{6} \left(\omega +ikt\right)^{3} -\frac{\theta _{2} }{12} \left(\omega +ikt\right)^{4} +\frac{i\theta _{3} }{20} \left(\omega +ikt\right)^{5} +\frac{\theta _{4} }{30} \left(\omega +ikt\right)^{6} \right]\right\}} , \end{equation}
\begin{equation} \label{4.38} Q_{0} \left(ikt\right)=\left\{\frac{1}{k^{2} } \left[-\frac{i\theta _{1} }{6} \left(ikt\right)^{3} -\frac{\theta _{2} }{12} \left(ikt\right)^{4} +\frac{i\theta _{3} }{20} \left(ikt\right)^{5} +\frac{\theta _{4} }{30} \left(ikt\right)^{6} \right]\right\} .\end{equation}
Calculation of the average $\left\langle x\right\rangle _{f} $ yields
\begin{equation} \label{4.39}  {\left\langle x\right\rangle _{f} =\frac{\int _{-\infty }^{+\infty }z\exp \left(zkt\right)f\left(z,0\right)dz }{\int _{-\infty }^{+\infty }\exp \left(zkt\right)f\left(z,0\right)dz } } {+\nu t+\left(\zeta k+\frac{\theta _{1} }{2} \right)t^{2} +\frac{\theta _{2} k}{3} t^{3} +\frac{\theta _{3} k^{2} }{4} t^{4} +\frac{\theta _{4} k^{3} }{5} t^{5} }. \end{equation}

For smooth correlator \eqref{4.2} $\theta _{1} =\sigma sq''\left(0\right)\le 0$; $\theta _{2} =\sigma ^{2} q\left(0\right)-s^{2} \left|q''\left(0\right)\right|$; $\theta _{3} \ge 0;\, \, \, \theta _{4} \ge 0$. Presuming again the Gaussian initial distribution, centered on some $x=X$, see \eqref{3.25}, we get
\begin{equation} \label{4.40}  {\left\langle x\right\rangle _{f} } {=X+\left(\frac{k}{2\alpha ^{2} } +\nu \right)t+\left(\zeta k-\frac{1}{2} \sigma s\left|q''\left(0\right)\right|\right)t^{2} +\frac{k}{3} \left(\sigma ^{2} q\left(0\right)-s^{2} \left|q''\left(0\right)\right|\right)t^{3} }  {+\frac{k^{2} }{2} s\sigma q\left(0\right)t^{4} +\frac{k^{3} }{5} s^{2} q\left(0\right)t^{5} }  .\end{equation}
Calculation of $\left\langle x^{2} \right\rangle _{f} $ yields
\begin{equation} \label{4.41)}  {\left\langle x\right\rangle _{f}^{2} =\left\{\frac{\int _{-\infty }^{+\infty }z\exp \left(zkt\right)f\left(z,0\right)dz }{\int _{-\infty }^{+\infty }\exp \left(zkt\right)f\left(z,0\right)dz } +\left[\left(\nu t+\zeta kt^{2} \right)+i\left. \frac{\partial }{\partial \omega } G\left(\omega ,t\right)\right|_{\omega =0} \right]\right\}^{2} } . 
\end{equation}
And for dispersion $\left\langle x^{2} \right\rangle _{f} -\left\langle x\right\rangle _{f}^{2} $ we get
\begin{equation} \label{4.42} {\left\langle x^{2} \right\rangle _{f} -\left\langle x\right\rangle _{f}^{2} =\frac{\int _{-\infty }^{+\infty }z^{2} \exp \left(zkt\right)f\left(z,0\right)dz }{\int _{-\infty }^{+\infty }\exp \left(zkt\right)f\left(z,0\right)dz } -\left[\frac{\int _{-\infty }^{+\infty }z\exp \left(zkt\right)f\left(z,0\right)dz }{\int _{-\infty }^{+\infty }\exp \left(zkt\right)f\left(z,0\right)dz } \right]^{2} }  {+2\zeta t+\theta _{2} t^{2} +\theta _{3} kt^{3} +\theta _{4} k^{2} t^{4} } .\end{equation}
Again for the Gaussian initial distribution, centered on some $x=X$, see \eqref{3.25}, we get
\begin{equation} \label{4.43}  {\left\langle x^{2} \right\rangle _{f} -\left\langle x\right\rangle _{f}^{2} =\frac{1}{2\alpha ^{2} } +2\zeta t} {+\left[\sigma ^{2} q\left(0\right)-s^{2} \left|q''\left(0\right)\right|\right]t^{2} +2s\sigma q\left(0\right)kt^{3} +s^{2} q\left(0\right)k^{2} t^{4} }. \end{equation}
In Section 6 we will discuss above expressions in some detail.

\section{The nonlinear autocatalysis}\label{Sec:5}
\setcounter{equation}{0}

It was shown in the previous Sections that for three different assumptions about the transfer mechanisms the average productivity is growing asymptotically, i.e. the objects, capable to self-instructed formation appear in the system inevitable. However, as it was shown, for the linear autocatalysis the dispersion of distribution is also growing asymptotically, i.e. there is no ``selection''. So the natural conclusion is: to obtain ``selection'' we, most probably, need to introduce the nonlinearity in kinetics. The model \eqref{3.10}-\eqref{3.11} is, therefore, modified:
\begin{equation} \label{5.1} \frac{\partial f\left(x,t\right)}{\partial t} =k\left[x-\gamma \left(t\right)\right]f^{n+1} \left(x,t\right)-\nu \frac{\partial f\left(x,t\right)}{\partial x} +\zeta \frac{\partial ^{2} f\left(x,t\right)}{\partial x^{2} } , \end{equation}
\begin{equation} \label{5.2} \gamma \left(t\right)=\frac{\int _{-\infty }^{+\infty }zf^{n+1} \left(z,t\right)dz }{\int _{-\infty }^{+\infty }f^{n+1} \left(z,t\right)dz } . \end{equation}
As usual we keep the condition \eqref{1.1}:
\begin{equation} \label{5.3} \int _{-\infty }^{+\infty }f\left(z,t\right)dz=1  .  \end{equation}
There is no general method for solving \eqref{5.1} exactly. Even more, as it will be evident below, the properties of solution and the proper approximations depend critically not on the power of kinetic term $n$ only, but on the functional form of the initial distribution $f\left(x,0\right)$ as well. To get some insight we consider here only a simplified model, presuming $\zeta \ll 1$ and dropping the diffusional term, i.e.
\begin{equation} \label{5.4} \frac{\partial f\left(x,t\right)}{\partial t} =k\left[x-\gamma \left(t\right)\right]f^{n+1} \left(x,t\right)-\nu \frac{\partial f\left(x,t\right)}{\partial x}  .  \end{equation}
For further comparison let us first consider the case of linear kinetics, $n=0$ (still, the problem as a whole is nonlinear, see \eqref{5.2}). The system, corresponding to this first-order partial differential equation is:
\begin{equation} \label{5.5)} dt=\frac{dx}{\nu } =\frac{df}{k\left[x-\gamma \left(t\right)\right]f}  .  \end{equation}
The first integral is $c_{1} =x-\nu t;\, \, \, x=c_{1} +\nu t$; and the second
\begin{equation} \label{5.6)} c_{2} =f\exp \left\{-k\left[c_{1} t+\frac{\nu }{2} t^{2} -\int _{}^{t}\gamma \left(t'\right)dt' \right]\right\} .  \end{equation}
Then the solution is
\begin{eqnarray} \label{5.7)}  & &{f=p\left(x-\nu t\right)\exp \left\{k\left[\left(x-\nu t\right)t+\frac{\nu }{2} t^{2} -\Gamma \left(t\right)\right]\right\};} \nonumber \\ & &{\Gamma \left(t\right)=\int _{}^{t}\gamma \left(t'\right)dt' } . \end{eqnarray}
Here $p\left(x-\nu t\right)$ is an arbitrary function. From the initial condition $f\left(x,0\right)=f_{0} \left(x\right)$ it follows $f_{0} \left(x\right)=p\left(x\right)$, and the solution is
\begin{equation} \label{5.8)} f=f_{0} \left(x-\nu t\right)\exp \left\{k\left[\left(x-\nu t\right)t+\frac{\nu }{2} t^{2} -\Gamma \left(t\right)\right]\right\} .  \end{equation}

To find $\Gamma \left(t\right)$ we use condition \eqref{5.3}:
\begin{equation} \label{5.9} \int _{-\infty }^{+\infty }f_{0} \left(z\right)\exp \left\{zkt\right\}dz= \exp \left\{k\left[-\frac{\nu }{2} t^{2} +\Gamma \left(t\right)\right]\right\} .  \end{equation}
Taking again initial distribution \eqref{3.25} as an example,
\begin{equation} \label{5.10} f_{0} \left(x\right)=\frac{\alpha }{\sqrt{\pi } } \exp \left[-\alpha ^{2} \left(x-X\right)^{2} \right] ,  \end{equation}
we obtain from \eqref{5.9}
\begin{equation} \label{5.11)} -\frac{\nu }{2} t^{2} +\Gamma \left(t\right)=Xt+k\left(\frac{t}{2\alpha } \right)^{2}  ,  \end{equation}
\begin{equation} \label{5.12} f\left(x,t\right)=\frac{\alpha }{\sqrt{\pi } } \exp \left\{-\alpha ^{2} \left[x-\left(\nu +\frac{k}{2\alpha ^{2} } \right)t-X\right]^{2} \right\} .  \end{equation}

So the initial distribution advances keeping its form. Remarkably, it advances even for zero or negative drift, $-\frac{k}{2\alpha ^{2} } <\nu \le 0$. Evidently this is due to the growth of the infinite ``positive tail'' of the peaked initial distribution. It is interesting to consider an ``opposite extreme'': a ``flat'' distribution, which occupies only limited domain in $x$ initially. Let us consider the ``rectangular'' initial distribution
\begin{equation} \label{5.13} f\left(x,0\right)=h\left\{{\rm H} \left(a_{2} -x\right)-{\rm H} \left(a_{1} -x\right)\right\};\, \, \, h=\frac{1}{a_{2} -a_{1} } . \end{equation}
Here ${\rm H} \left(y\right)$ is the Heaviside function; we presume ${\rm H} \left(0\right)=0$. Then \eqref{5.9} yields
\begin{equation} \label{5.14)} \exp \left\{k\left[-\frac{\nu }{2} t^{2} +\Gamma \left(t\right)\right]\right\}=h\int _{a_{1} }^{a_{2} }\exp \left(zkt\right)dz =\frac{h}{kt} \left[\exp \left(a_{2} kt\right)-\exp \left(a_{1} kt\right)\right].   \end{equation}
Correspondingly, the solution is
\begin{equation} \label{5.15} f\left(x,t\right)=\frac{kt\exp \left[kt\left(x-\nu t\right)\right]\left[{\rm H} \left(a_{2} -\left(x-\nu t\right)\right)-{\rm H} \left(a_{1} -\left(x-\nu t\right)\right)\right]}{\left[\exp \left(a_{2} kt\right)-\exp \left(a_{1} kt\right)\right]} .   \end{equation}

Evidently, the initially occupied domain  $a_{1} +\nu t\le x\le a_{2} +\nu t$ moves with constant drift velocity; however the initial rectangle transforms into permanently sharpening asymmetric ``tooth'', with asymptotically linearly growing ``front edge'' and nearly exponentially decaying ``rear edge'':
\begin{equation} \label{5.16} f\left(a_{2} +\nu t-\varepsilon ,t\right)=\frac{kt}{1-\exp \left[-\left(a_{2} -a_{1} \right)kt\right]} ;\, \, \varepsilon \to +0 ,  \end{equation}
\begin{equation} \label{5.17} f\left(a_{1} +\nu t,t\right)=\frac{kt\exp \left[-\left(a_{2} -a_{1} \right)kt\right]}{1-\exp \left[-\left(a_{2} -a_{1} \right)kt\right]}  .  \end{equation}
Now, let us consider the nonlinear kinetics, i.e. $n\ge 1$ in \eqref{5.1} and \eqref{5.2}. The corresponding system is now
\begin{equation} \label{5.18)} dt=\frac{dx}{\nu } =\frac{df}{k\left[x-\gamma \left(t\right)\right]f^{n+1} }  .  \end{equation}
The first integrals are now
\begin{eqnarray} \label{5.19)} & & {c_{1} =x-\nu t}, \nonumber \\ & &{c_{2} =\frac{1}{nf^{n} } +k\left[c_{1} t+\frac{\nu }{2} t^{2} -\Gamma \left(t\right)\right]\, ;}  \end{eqnarray}
and the solution is
\begin{equation} \label{5.20)} \frac{1}{nf^{n} } +k\left[\left(x-\nu t\right)t+\frac{\nu }{2} t^{2} -\Gamma \left(t\right)\right]=p\left(x-\nu t\right) .  \end{equation}
Here $p\left(y\right)$ is an arbitrary function, which is determined by the initial condition $f\left(x,0\right)=f_{0} \left(x\right)$, yielding
\begin{eqnarray}\label{5.21} & &\frac{1}{f\left(x,t\right)^{n} } =\frac{1}{f_{0} \left(x-\nu t\right)^{n} } -nk\left[\left(x-\nu t\right)t+\frac{\nu }{2} t^{2} -\Gamma \left(t\right)\right], \nonumber  \\
& &f\left(x,t\right)=\frac{1}{\left\{\frac{1}{f_{0} \left(x-\nu t\right)^{n} } -nk\left[\left(x-\nu t\right)t+\frac{\nu }{2} t^{2} -\Gamma \left(t\right)\right]\right\}^{\frac{1}{n} } } . \end{eqnarray}
Constraint \eqref{5.3} yields equation for finding $\Gamma \left(t\right)$:
\begin{equation} \label{5.22} \int _{-\infty }^{+\infty }\frac{dz}{\left\{\frac{1}{f_{0} \left(z\right)^{n} } -nk\left[zt+\frac{\nu }{2} t^{2} -\Gamma \left(t\right)\right]\right\}^{\frac{1}{n} } }  =1 .  \end{equation}

Evidently, it is impossible to resolve this equation for arbitrary $f_{0} \left(z\right)$ and $n$. So we take $n=1$ and two forms of initial distribution. We take first
\begin{equation} \label{5.23} f_{0} \left(x\right)=\frac{\alpha }{\pi \left[\alpha ^{2} \left(x-X\right)^{2} +1\right]}  .  \end{equation}
This is the famous Cauchy-Lorentz distribution, see \cite{6}, p.~51. For $\alpha \to \infty \, \, \, \, \, f_{0} \left(x\right)\to \delta \left(x-X\right)$, i.e. for large $\alpha $ this function is very close to \eqref{5.10}. Substitution of \eqref{5.23} into \eqref{5.21} and \eqref{5.22} for $f_{0} \left(z\right)$ yields
\begin{equation} \label{5.24)}  {f\left(x,t\right)} {=\left\{\pi \alpha \left(x-\nu t-X\right)^{2} -k\left(x-\nu t-X\right)t-k\left[-\frac{\pi }{\alpha k} +Xt+\frac{\nu }{2} t^{2} -\Gamma \left(t\right)\right]\right\}^{-1} } , \end{equation}
\begin{equation} \label{5.25} \int _{-\infty }^{+\infty }\frac{dz}{\alpha \pi z^{2} -ktz-k\left[Xt+\frac{\nu }{2} t^{2} -\frac{\pi }{\alpha k} -\Gamma \left(t\right)\right]}  =1 .  \end{equation}
Denoting for convenience,
\begin{eqnarray} \label{5.26)}  & &{a=\pi \alpha , \, \, \, b=-kt, \, \, \, }\nonumber \\ & &{c=-k\left[Xt+\frac{\nu }{2} t^{2} -\frac{\pi }{\alpha k} -\Gamma \left(t\right)\right]},  \end{eqnarray}
we rewrite \eqref{5.25} as
\begin{equation} \label{5.27)} \int _{-\infty }^{+\infty }\frac{dz}{az^{2} +bz+c}  =1 .  \end{equation}
The quadratic polynomial in denominator should not have real roots, i.e. $b^{2} <4ac$. For this case, see e.g. \cite{7}, the integral is
\begin{equation} \label{5.28)} \int _{-\infty }^{+\infty }\frac{dz}{az^{2} +bz+c}  =\frac{2}{\sqrt{4ac-b^{2} } } \left. \arctan \frac{2az+b}{\sqrt{4ac-b^{2} } } \right|_{-\infty }^{+\infty } =\frac{2\pi }{\sqrt{4ac-b^{2} } } .   \end{equation}
Correspondingly $4ac-b^{2} =4\pi ^{2} $, i.e.
\begin{eqnarray} \label{5.29)}  -& &4\pi \alpha k\left[Xt+\frac{\nu }{2} t^{2} -\frac{\pi }{\alpha k} -\Gamma \left(t\right)\right]-k^{2} t^{2} =4\pi ^{2} , \nonumber \\ & &{Xt+\frac{\nu }{2} t^{2} -\Gamma \left(t\right)=-\frac{kt^{2} }{4\pi \alpha } } . \end{eqnarray}
And the solution is
\begin{equation} \label{5.30} f\left(x,t\right)=\frac{1}{\frac{\pi }{\alpha } \left[\alpha ^{2} \left(x-\left(\nu +\frac{k}{2\pi \alpha } \right)t-X\right)^{2} +1\right]}  .  \end{equation}

Remarkably, the initial distribution \eqref{5.23} advances keeping its form even for strongly nonlinear quadratic autocatalysis. And, as in the linear autocatalysis case it advances even for zero- or negative drift. Again this is due to the growth of the infinite ``positive tail'' of the peaked initial distribution. So now we consider an ``opposite extreme'', i.e. the ``rectangular'' initial distribution \eqref{5.13}. For this initial distribution and $n=1$ the condition \eqref{5.22} becomes
\begin{equation} \label{5.31)} h\int _{a_{1} }^{a_{2} }\frac{dz}{\left[1-kh\left(\frac{\nu }{2} t^{2} -\Gamma \left(t\right)\right)\right]-\left(kht\right)z}  =1 .  \end{equation}
Denoting for convenience,
\begin{equation} \label{5.32)} w=\left[1-kh\left(\frac{\nu }{2} t^{2} -\Gamma \left(t\right)\right)\right];\, \, \, \, g=\left(kht\right) ,  \end{equation}
we have
\begin{equation} \label{5.33)} \int _{a_{1} }^{a_{2} }\frac{dz}{w-gz}  =\frac{1}{h} ;\, \, \,  \end{equation}
then
\begin{equation} \label{5.34)} w=g\frac{\left(a_{2} -a_{1} \exp \left(-\frac{g}{h} \right)\right)}{\left(1-\exp \left(-\frac{g}{h} \right)\right)} ,\, \, \, \,  \end{equation}
\begin{equation} \label{5.35)} \frac{\nu }{2} t^{2} -\Gamma \left(t\right)=\frac{1}{kh} -t\frac{a_{2} -a_{1} \exp \left(-kt\right)}{1-\exp \left(-kt\right)}  .  \end{equation}
And the solution is
\begin{equation} \label{5.36)} f\left(x,t\right)=\frac{{\rm H} \left(a_{2} -\left(x-\nu t\right)\right)-{\rm H} \left(a_{1} -\left(x-\nu t\right)\right)}{kt\left[\frac{a_{2} -a_{1} \exp \left(-kt\right)}{1-\exp \left(-kt\right)} -\left(x-\nu t\right)\right]}  .  \end{equation}
Similar to \eqref{5.16}, \eqref{5.17}, the initially occupied domain $a_{1} +\nu t\le x\le a_{2} +\nu t$ moves with constant drift velocity; the initial rectangle transforms into permanently sharpening asymmetric ``tooth'', with nearly exponentially growing ``front edge'' and decaying ``rear edge''.

\section{Discussion}\label{Sec:6}
\setcounter{equation}{0}

Modeling the emergence of self-constructing objects in the open system, we considered the continuous master equation \eqref{1.2} with rather general assumptions about the kernel of the integral ``transition term''. Following \cite{3}, in Section 1 we used the difference kernel, $w\left(x-x'\right)$. We obtained the exact solution (by quadrature) of the nonlinear equation \eqref{1.2} for arbitrary functional form of the $w\left(z\right)$ and for arbitrary initial distribution $f\left(x,0\right)$. For this general case the mean productivity $\left\langle x\right\rangle $ and dispersion $\left\langle x^{2} \right\rangle -\left\langle x\right\rangle ^{2} $ of the distribution function $f\left(x,t\right)$ are calculated, see \eqref{2.15}, \eqref{2.18}. The main result, as it was formulated in \cite{3}, is: even if the initial distribution lies completely in the non-self-constructing domain, the average value of productivity $\left\langle x\right\rangle $ becomes positive for large enough time. It means that even if the self-constructing objects are absent in a particular system, sooner or later they appear inevitably, if their formation is generally possible via a chain of small random changes initiated by the environment.

We also studied the time dependence of the mean productivity and dispersion; the dispersion is growing fast; i.e. there is no ``selection of the best'', the distribution flattens asymptotically. As an example we calculated the mean productivity and dispersion for the Gaussian kernel \eqref{2.21} and for the Gaussian initial distribution \eqref{2.22}, see \eqref{2.23} and \eqref{2.24}. From expression \eqref{2.23} it is evident, that for the Gaussian form of both $w\left(z\right)$ and $f\left(x,t\right)$ with finite $\alpha $ and $\beta $ there are two different mechanisms of $\left\langle x\right\rangle $ growth. For small $\alpha $ (wide initial distribution) and large $\beta $ (short-range transition probability) the evolution is determined by the width of the initial distribution and reaction kinetics, and is initially nearly linear in time. For this case dispersion \eqref{2.24} is also growing nearly linear initially. On the other hand, for large $\alpha $ (narrow initial distribution), and small $\beta $ (wide-range transition probability) the growth is determined essentially by the transition range and reaction kinetics and is nearly exponential from the very beginning. The dispersion is growing even faster than exponential. Naturally, the second mechanism is always dominating asymptotically. According to \eqref{2.24} dispersion is growing monotonically. The only case when the dispersion remains constant and $\left\langle x\right\rangle $ is growing exactly linear in time is the limit $\beta \to \infty $, i.e. when $w\left(z\right)$ becomes the $\delta $-function; however this limit looks not quite reasonable physically.

In Section 3, following the well known procedure \cite{4}, we developed the diffusional approximation \eqref{3.11} for the master equation \eqref{1.2}. The equation similar to \eqref{3.11} was introduced \textit{ad hoc} in \cite{8} as a Fitness-Space model\textit{ }of the RNA Virus Evolution; however only approximate and numerical solutions were given. We obtained exact solution of \eqref{3.11} for the arbitrary initial distribution, see \eqref{3.30}. Remarkably, for the $\delta $-functional initial distribution we get the solution \eqref{3.31}, which looks very similar to the well known solution of the diffusion equation with a drift; however with linearly increasing effective ``drift velocity''. For the general initial distribution we calculated the mean productivity $\left\langle x\right\rangle $ and dispersion $\left\langle x^{2} \right\rangle -\left\langle x\right\rangle ^{2} $, see \eqref{3.22} and \eqref{3.24}; both of them are increasing asymptotically. For the Gaussian initial distribution \eqref{3.25} they become simple polynomial expressions. Interestingly, while for the negative drift, $\nu <0$ (preferable ``backwards jumps''), the productivity may change non-monotonically, i.e. decrease till some time $t_{s} $, see \eqref{3.28}, the dispersion always increases monotonically.

Considering the master equation in Section 3 we have, following \cite{4}, neglected the dependence of the jump probability moments \eqref{3.7} on $x$, i.e. on the starting point of the jump. To study the ``opposite extreme'' in Section 4 we presumed this dependence to be a random function of $x$. Starting again with equation \eqref{3.8}, we use the moments $\phi $ and $\psi $ given by \eqref{4.1}, i.e. depending on the Gaussian random function $\eta \left(x\right)$. Then \eqref{3.8} takes the form \eqref{4.3}. Averaging this equation over fluctuations of $\eta \left(x\right)$ and using the the Novikov-Furutsu formula \cite{5} yields \eqref{4.15}; this is the fourth order equation, and the coefficients at the $x$-derivatives become time-dependent. Fourier transform of the exact solution of \eqref{4.15} is given by \eqref{4.34}-\eqref{4.38}. It is impossible to perform the inverse transform for arbitrary initial distribution; however the mean productivity and dispersion could be calculated, see \eqref{4.39} and \eqref{4.42}. For the Gaussian initial distribution we get explicit expressions \eqref{4.40} and \eqref{4.43} for the mean productivity and dispersion, respectively. The randomness results in the additional polynomial in the time dependence of both mean productivity and dispersion, with the polynomial coefficients depending on the functional forms of the transition probability and correlator. These expressions are still too complicated. To get some physical insight, let us consider two limiting cases. First, let us presume the random addition in the diffusion term to be negligible. Then expressions \eqref{4.40} and \eqref{4.43} simplify to

\begin{equation} \label{6.1)} \left\langle x\right\rangle _{f} =X+\left(\frac{k}{2\alpha ^{2} } +\nu \right)t+\zeta kt^{2} +\frac{k}{3} \left(\sigma ^{2} q\left(0\right)\right)t^{3}  .  \end{equation}

\begin{equation} \label{6.2)} \left\langle x^{2} \right\rangle _{f} -\left\langle x\right\rangle _{f}^{2} =\frac{1}{2\alpha ^{2} } +2\zeta t+\left[\sigma ^{2} q\left(0\right)\right]t^{2}  .  \end{equation}

I.e. as in the constant coefficient case, see Section 3, the non-monotonous growth of the average productivity is possible for the negative constant drift, $\nu <0$, only; the dispersion is always growing monotonically.

On the other hand, if the random addition in the drift term is negligible, we have

\begin{equation} \label{6.3)} \left\langle x\right\rangle _{f} =X+\left(\frac{k}{2\alpha ^{2} } +\nu \right)t+\zeta kt^{2} -\frac{k}{3} \left(s^{2} \left|q''\left(0\right)\right|\right)t^{3} +\frac{k^{3} }{5} s^{2} q\left(0\right)t^{5} , \end{equation}

\begin{equation} \label{6.4)} \left\langle x^{2} \right\rangle _{f} -\left\langle x\right\rangle _{f}^{2} =\frac{1}{2\alpha ^{2} } +2\zeta t-\left[s^{2} \left|q''\left(0\right)\right|\right]t^{2} +s^{2} q\left(0\right)k^{2} t^{4} . \end{equation}

Evidently, if the fluctuations are strong enough and/or the constant diffusion coefficient is small, both the average productivity (even for $\nu >0$) and dispersion can grow non-monotonically for some intermediate times. However, asymptotically they both grow even faster, than in the former case. In the sense of ``selection'' this case is even worse, the distribution function flattens much faster; generally, the randomness in the diffusive term influences the whole process essentially stronger, than the randomness in the drift term.

In Sections 2-4 we explored the behavior of our model for three different assumptions about the transfer mechanism, i.e. the way of random change of the productivities. For all assumptions two conclusions are common: the average productivity is growing asymptotically, i.e. the objects, capable to self-instructed formation appear in the system inevitable; the dispersion of distribution function is also growing asymptotically, i.e. there is no ``selection of the best''. So we can conclude that the last feature is due to linear self-catalysis; therefore it is interesting to study the non-linear kinetics.

The model with non-linear self-catalysis is defined by \eqref{5.1}-\eqref{5.3}. We could not solve this system exactly. As it follows from the consideration in Section 5, the properties of solution and the proper approximations depend critically not on the power of kinetic term $n$ only, but on the functional form of the initial distribution $f\left(x,0\right)$ as well. So, postponing the approximate solution of the system \eqref{5.1}-\eqref{5.3} to the further communications, presuming $\zeta \ll 1$ we considered the system without diffusion. The resulting system \eqref{5.2}-\eqref{5.4} is again exactly solvable, at least for some particular forms of the initial distribution and quadratic kinetics. To compare with the quadratic case we first obtained exact solution of \eqref{5.4} for $n=0$ and two forms of the initial distribution: the Gaussian form \eqref{3.25} and the ``rectangular'' distribution \eqref{5.13}. For the Gaussian form the initial distribution advances keeping its form, see \eqref{5.12}, even for zero or negative drift (which is due to the infinite positive ``tail''). On the other hand, for the rectangular distribution the initially occupied domain moves with the drift velocity $\nu $ \eqref{5.15}; the initial rectangle transforms into permanently sharpening asymmetric ``tooth'', with growing ``front edge'' and decaying ``rear edge'', see \eqref{5.16}-\eqref{5.17}.

For the quadratic autocatalysis, $n=1$, we again considered two initial distributions: the Cauchy-Lorentz distribution \eqref{5.23} and the rectangular distribution \eqref{5.13}. The exact solution for the Cauchy-Lorentz initial distribution is given by \eqref{5.30}; remarkably the initial distribution advances preserving its form. Again due to infinite positive ``tail'' the distribution advances even for zero or negative $\nu $. The positive addition to the drift velocity is proportional to $k$ and inverse proportional to $\alpha $; it increases with flattening of the initial distribution. This positive addition disappears in the limit $\alpha \to \infty $, when the Cauchy-Lorentz distribution approaches $\delta $-function. For the rectangular distribution \eqref{5.13} the initially occupied domain moves with the drift velocity $\nu $; again the rectangular distribution transforms into permanently sharpening asymmetric ``tooth''. Interesting difference with the linear autocatalysis is: for the linear case the ``front edge'' is asymptotically growing linearly, while for the quadratic case it is growing nearly exponentially. The ``rear edge'' is decaying nearly exponentially in the linear case and as $t^{-1} $ in the quadratic case; however  the decay rate for the linear case was dependent both on the size of initially occupied domain $a_{2} -a_{1} $ and kinetic coefficient $k$, while for the quadratic case the growth rate depends on $k$ only. I.e., while the size of initially occupied domain is preserved, for the latter case the asymptotic evolution of the distribution's form is not dependent on its initial state, whether ``narrow'' or ``wide'' the initial rectangle was. For the sharply peaked initial distribution with the infinite positive ``tail'' both for linear and quadratic autocatalysis the initial distribution has advanced keeping its form. On the other hand, for the initial distribution limited to a finite domain in the quadratic case the advantage of the objects with higher productivity is more pronounced. These properties of the exact solutions of \eqref{5.4} should be taken into account in the construction of approximate solutions of \eqref{5.1}-\eqref{5.3} for $n\ge 1$.

Summing up, our model may be considered as an illustration of a general question: suppose a system is capable to self-organization of some kind if a governing parameter reaches some ``threshold value''. Now, let us consider an ensemble of such systems initially in non-self-organized state, under external limitation on their total number. Then let us presume additionally that the governing parameter of each system is randomly changing. The question is: will the probability to find self-organized system increase? In other words: will the systems with the governing parameter accidentally above the threshold value ``survive'' or (due to external limitation) disappear from the ensemble? For the model, considered here, with the productivity as a governing parameter, and the self-constructing objects as ``self-organized'', the probability is increasing, the fraction of self-constructing objects is growing. However asymptotically fast growth of the distribution dispersion shows the absence of real advantage of the ``best self-organized objects''. On the other hand, the examples given in Section 5 suggest that for such advantage to exist the self-catalysis of the order higher than linear may be preferential.


{\small \topsep 0.6ex

}

\end{document}